\documentclass[a4paper]{jpconf}
\usepackage{graphicx}
\usepackage{amsfonts,amsmath,amssymb}
\usepackage{textcomp}
\usepackage{bm}
\usepackage{graphicx}
\usepackage[latin1]{inputenc}
\usepackage{array}
\usepackage{gensymb}
\usepackage{cite}
\renewcommand{\vec}[1]{\bm{#1}}
\newcommand{\ee}{\mathrm{e}}

\newcommand{\pp}{\partial^{{}}}

\newcommand{\fracsmall}[2]{\mbox{$\frac{#1}{#2}$}}

\newcommand{\AAA}{\vec{A}}

\newcommand{\GGn}{\vec{G}}

\newcommand{\HHH}{\vec{H}}

\newcommand{\RRR}{\vec{R}}

\newcommand{\uuu}{\vec{u}}

\newcommand{\eps}{\epsilon}

\newcommand{\beq}[1]{\begin{equation} \eqlab{#1}}
\newcommand{\eeq}{\end{equation}}
\newcommand{\bsub}{\begin{subequations}}
\newcommand{\esub}{\end{subequations}}
\def\bal#1\eal{\begin{align}#1\end{align}}
\def\bsubal#1\esubal{\bsub \begin{align}#1\end{align} \esub}

\newcommand{\eqlab}[1]{\label{eq:#1}}
\renewcommand{\eqref}[1]{Eq.~(\ref{eq:#1})}

\newcommand{\figref}[1]{Fig.~\ref{fig:#1}}

\begin{document}
 \title{Bubbles in graphene - a computational study}
 
 \author{Mikkel Settnes, Stephen R. Power, Jun Lin, Dirch H. Petersen and Antti-Pekka Jauho} 
 
 \address{Center for Nanostructured Graphene (CNG),\\ DTU Nanotech, Technical University of Denmark, DK-2800 Kgs. Lyngby, Denmark}

\ead{Antti-Pekka.Jauho@nanotech.dtu.dk}

\begin{abstract}
Strain-induced deformations in graphene are predicted to give rise to large pseudomagnetic fields.
We examine theoretically the case of gas-inflated bubbles to determine whether signatures of such fields are present in the local density of states.
Sharp-edged bubbles are found to induce Friedel-type oscillations which can envelope pseudo-Landau level features in certain regions of the bubble.
However, bubbles which minimise interference effects are also unsuitable for pseudo-Landau level formation due to more spatially varying field profiles.
\end{abstract}


\section{Introduction}
Strain engineering has been proposed as a method to manipulate the electronic, optical and magnetic properties of graphene \cite{Guinea2009,Levy2010,Lu2012,Jones2014, Zenan2014,Low2011,Pereira2009,Neek-Amal2013,Power2012,Pereira2009prl}.
It is based on the close relation between the structural and electronic properties of graphene.
An inhomogeneous strain field can introduce pseudomagnetic fields (PMFs),\cite{Guinea2009,Zenan2014,Jones2014}
where the altered tight binding hoppings mimic the role of a gauge field in the low energy effective Dirac model of graphene \cite{Suzuura2002,Vozmediano2010}.
Guinea \etal \cite{Guinea2009} demonstrated that nearly homogeneous PMFs can be generated by applying triaxial strain.
One of the most striking consequences of homogeneous PMFs is the appearance of a Landau-like quantization. \cite{Guinea2009,Neek-Amal2013}
Scanning tunnelling spectroscopy on bubble-like deformations has observed pseudo-Landau levels corresponding to PMFs stronger than 300 T \cite{Lu2012,Levy2010}.

Deformations can be induced in graphene samples by different techniques like pressurizing suspended graphene \cite{Bunch2008,Zenan2014} or by exploiting the thermal expansion coefficients of different substrates \cite{Lu2012}.
As a result, introducing nonuniform strain distributions at the nanoscale is a promising route towards strain engineering.
The standard theoretical approach to treat strain effects employs continuum mechanics to obtain the strain field.
The strain field can then be coupled to an effective Dirac model of graphene to study the generation of PMFs in various geometries.
In most studies, only the PMF distribution is considered as opposed to experimentally observable quantities like local density of states (LDOS).
This study calculates the LDOS of such systems without applying periodicity, which can introduce spurious interactions between neighboring bubbles.

\section{Model}

\subsection{Patched Green's function approach}
The patched Green's function  approach, developed in Ref \cite{Settnes2015},  treats device `patches' embedded within an extended two dimensional system described by a tight-binding Hamiltonian. 
This approach allows us to insert a single bubble into an otherwise pristine infinite graphene sheet, and avoids issues such as interferences between a bubble and its periodic images or system edges. 
The extended part of the system is treated through a self-energy $\Sigma_B$ entering the device area. 
\bal
\GGn_{D,D} &= \big(E\mathbf{1}-\HHH_{D,D}- \mathbf{\Sigma}_B \big)^{-1}  ,
\eal
The self-energy is written in terms of Green's functions (GFs) of an infinite, pristine sheet so that it can be calculated using methods taking advantage of periodicity and analytic integrability \cite{Power2011}. 
The Hamiltonian for the device region can be tridiagonalized allowing the GF of the device region, $\GGn_{D,D}$, to be treated using an adaptive recursive method.
A dual recursive sweep allows for efficient calculation of local properties everywhere in the device region surrounding a bubble, enabling us to investigate spatial variations of the LDOS.

\begin{figure}
	\centering
	\includegraphics[width=0.95\columnwidth]{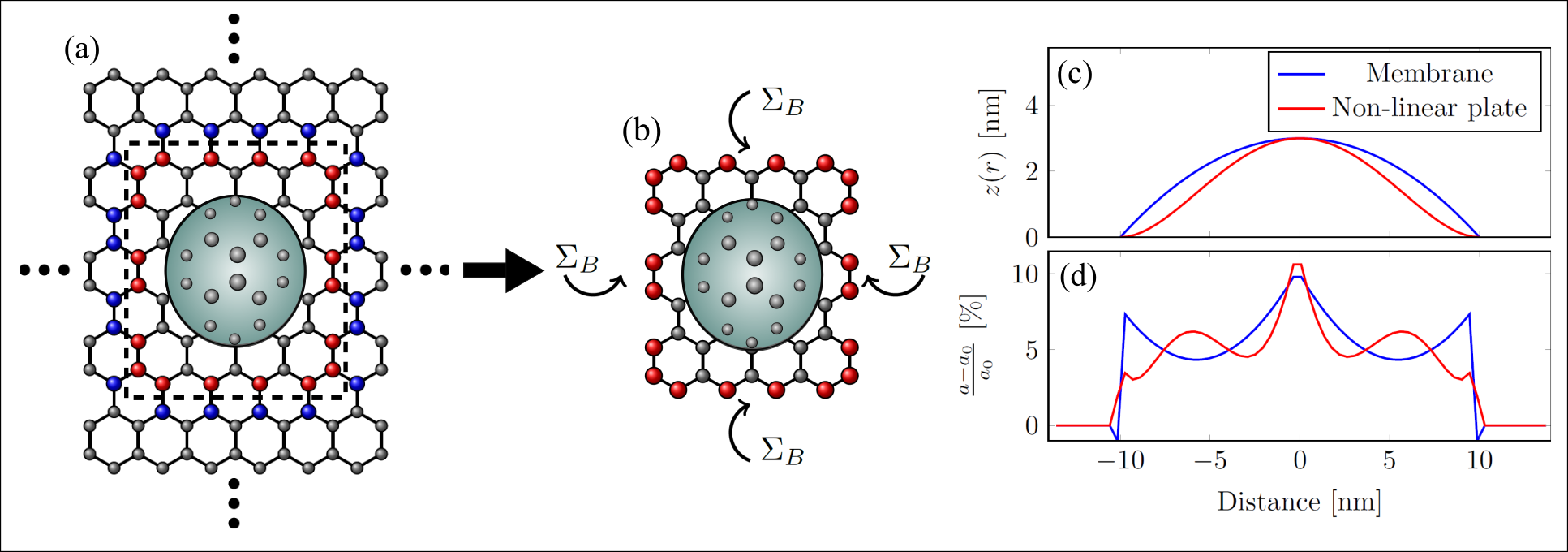}
	\caption[]{\small The patched GF method describes the extended graphene sheet away from the bubble device region (a) with a self energy term $\mathbf{\Sigma}_B$ (b). We consider membrane and non-linear plate type bubbles with radially dependent (c) height and (d)strain profiles.
	} \label{fig:PMF}
\end{figure}

\subsection{Strain model}
In this work we consider two possibilities for the shape of a gas inflated bubble, namely the \emph{membrane} and \emph{non-linear plate} models.\cite{Yue2012} 
The membrane model is suitable for very large bubbles where bending stiffness can be neglected, whereas the non-linear plate model is more appropriate for including bending effects near the edges of smaller bubbles.
Membrane bubbles therefore have very sharp edges, whereas the edges are smoother in the non-linear plate bubbles.
While these continuum models have been found to agree well with experimental shape profiles, more accurate modeling of bubble shapes and strain distributions can be achieved using molecular dynamics simulations. \cite{Neek-Amal2012a,Zenan2014,Qi2013}

Using the deformation field $\uuu = (u_r,u_\theta,z)$ the position of the atom $i$ initially at $\RRR^0_i$ becomes $\RRR_i = \RRR^0_i + \uuu$. The new bond lengths are afterwards determined as $d_{ij} = |\RRR_i - \RRR_j|$. Hoppings are modified according to
$
t_{ij}= t_0 \ee^{-\beta\big(d_{ij} /a_0 -1\big)},
$
where $t_0 \approx 2.7$ eV is the pristine coupling, $a_0 = 1.42\mathrm{\AA}$ is the carbon-carbon distance and $\beta \approx \pp \log(t)/\pp \log(a) |_{a=a_0} \approx 3.37$ \cite{Pereira2009}.
The strain tensor created by the strain field, $\eps_{ij} = \fracsmall{1}{2}\big( \pp_j u_i + \pp_i u_j + \pp_i z \pp_j z\big)$, gives rise to a gauge field \cite{Suzuura2002},
\bal 
\AAA = -\frac{\hbar\beta}{2ea_0}  \begin{pmatrix}
	\eps_{xx}-\eps_{yy} \\ 2 \eps_{xy}
\end{pmatrix},
\eal
with the resulting PMF given by $\mathbf{B_s} = \nabla \times \mathbf{A}$.
Table \ref{pseudotable} shows the height profile, in-plane displacements and PMFs for the two types of bubbles considered here.

\begin{table}
	\centering

	\begin{tabular}{| c | c | c | c | }
		\hline
		Model type &  $z(r,\theta)$ & $\uuu(r,\theta) = \begin{pmatrix}u_r \\u_\theta \end{pmatrix} $ &  $B_s (r,\theta)$  \\
		\hline
		Membrane & $h_0 \bigg( 1- \frac{r^2}{R^2}\bigg)$ & $\begin{pmatrix}	u_0 \frac{r}{R} \bigg(1-\frac{r}{R}\bigg) \\0 \end{pmatrix} $ & $\frac{\hbar\beta u_0}{2e a_0R^2}  \sin(3\theta)$  \\
		\hline
		Non-linear plate & $h_0 \bigg( 1- \frac{r^2}{R^2}\bigg)^2$ & $ \begin{pmatrix}r(R-r)(c_1+c_2 r) \\0 \end{pmatrix}$ & $\begin{aligned}
		\frac{\hbar\beta}{2e a_0} & \bigg[ (c_1 -c_2R) \\ &- \frac{32 h_0^2 r^3}{R^6}\bigg(1-\frac{r^2}{R^2}\bigg)\bigg]\sin(3\theta)
		\end{aligned}$
		\\
		\hline
		
	\end{tabular}
	
	\caption{Height profile $z(r,\theta)$, in-plane displacements $\uuu(r,\theta)$ and PMF distributions $B_s (r,\theta)$ for the membrane and non-linear plate bubbles, where $R$ and $h_0$ are the bubble radius and height respectively, and $u_0 = 1.136 h_0^2/R$, $c_1 = 1.308 h_0^2/R^3$ and $c_1 = -1.931 h_0^2/R^4$. \cite{Yue2012} }
	\label{pseudotable}
\end{table}

\section{Results}
\begin{figure}
	\centering
	\includegraphics[width=0.95\columnwidth]{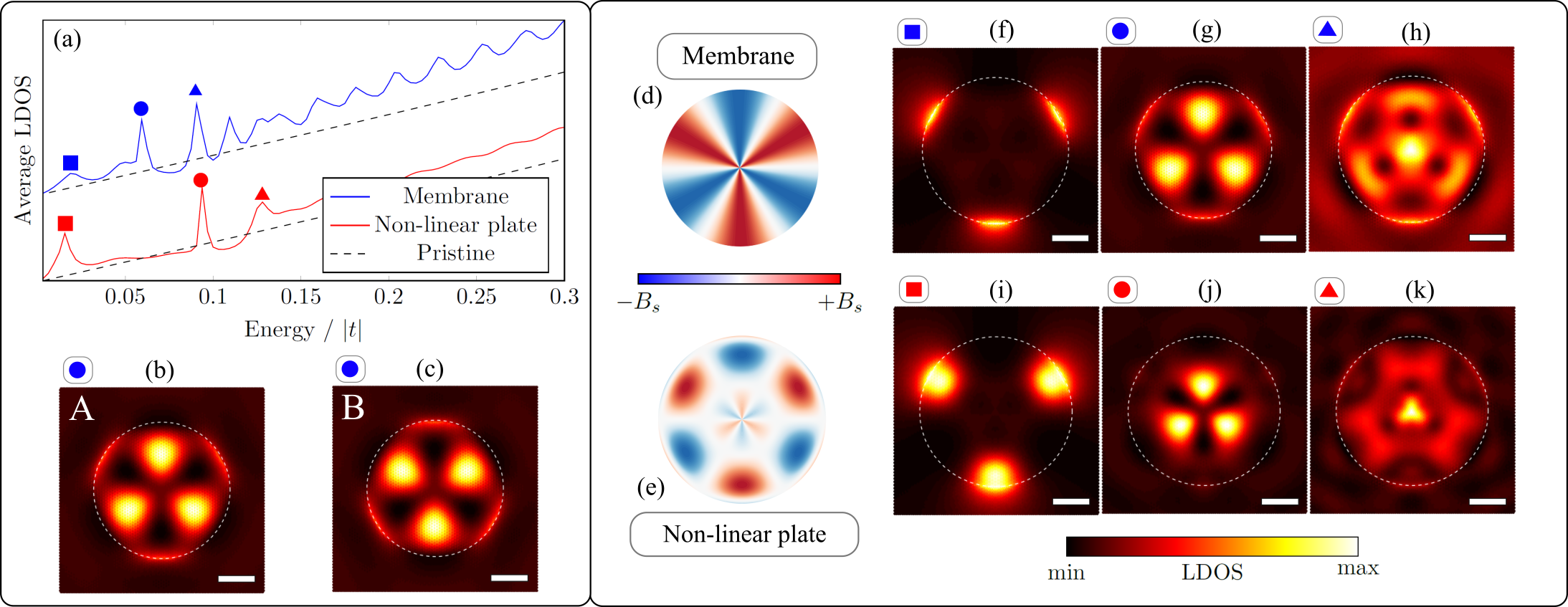}
	\caption[]{a) shows the averaged DOS with each bubble model with $R=10$nm and $h_0=10$nm. Important peaks in each are highlighted by symbols. b,c) The LDOS for the membrane model at the circle energy is mapped for the $A$ and $B$ sublattices separately. d,e) show the PMF distributions for each bubble type and f-k) show A sublattice LDOS maps for the peaks highlighted in a). The scale bar in all LDOS maps is $5$nm.
	} \label{fig:results}
\end{figure}

New hopping parameters, calculated from atomic coordinates generated by the continuum model displacements, give a tight-binding description of the bubble region which can be used within the patched GF approach. 
From this, the LDOS at every site in the bubble, as well as the average DOS, can be quickly calculated.
\figref{results}a shows the averaged DOS for membrane (blue) and non-linear plate (red) model bubbles. 
For the membrane model, we have previously distinguished between two different type of oscillations\cite{Settnes2015}.
A series of sharp peaks, such as those highlighted by the blue circle and triangle, were found to have an energy dependence, $E_n \sim \sqrt{n}$, consistent with Landau-like levels arising due to the PMF. 
In addition, the periodic oscillations visible at higher energies are identified as Friedel-type oscillations arising due to scattering of electrons induced by the sharp edges of the membrane-model bubble.

Before discussing the interplay between different oscillation types, we note that both features vary independently with position throughout the bubble region. 
The position dependence of the pseudo-Landau level features arises due to the non-uniform PMF distribution within the bubble, which is plotted in \figref{results}d. 
This takes maximum amplitudes along the armchair directions which occur every $60 \degree$, but is only three-fold symmetric due to a sign change between two consecutive amplitude maxima.
Unlike real magnetic fields, PMFs conserve time-reveral symmetry by taking opposite signs in the $K$ and $K'$ valleys of graphene.
One manifestation of this is a strong sublattice polarisation \cite{Settnes2015,Moldovan2013,Schneider2015}, which is clearly visible for the sublattice-split LDOS maps shown in \figref{results}b-c, where the circle LDOS peak from \figref{results}a is localized in different regions for different sublattices.
Comparison to panel \ref{pseudotable} confirm that these correspond to a change in the sign of the PMF. 
We note that this, the first ``pseudomagnetic peak", is localized along armchair directions where the PMFs are largest and reasonably constant.
The position dependence of the Friedel oscillations meanwhile emerges from interference between electrons scattered at different sides of the bubble.
The interplay of both oscillation types in membrane bubbles leads to the Friedel type acting as an envelope and quenching the LDOS signature of pseudo-Landau peaks in certain regions of the bubble, as is clear for the dark spots in the LDOS map for the higher energy in \figref{results}h.
STS measurements taken at such a spot would completely omit this peak due to the enveloping effects of the edge-induced Friedel oscillations.

When searching for signatures of PMFs in gas-inflated bubbles, it may thus be worth considering bubbles with a softer edge profile, such as the non-linear plate model, which should give riser to weaker Friedel oscillations. 
This is clear from the averaged LDOS curve in Fig. \ref{fig:results}a, where the higher energy oscillations are considerably suppressed compared to the membrane case.
However, in this case there is also an absence of sharp Landau-level-like peaks following a $\sqrt{n}$ distribution, with the possible exception of the peak denoted by the red circle.
This lack of pseudo-Landau features is consistent with the PMF distribution for this bubble type, plotted in \figref{results}e.
We note that this bubble, with less sharp edges, also has a radial fluctuation in the sign and strength of the PMF.
The center of the bubble has a field distribution similar to that of the membrane case, and the central region of the LDOS map in Fig \ref{fig:results}j resembles that of the corresponding membrane model peak (Fig \ref{fig:results}g). 
We note also that the Friedel features for the higher (triangle) energy in Fig. \ref{fig:results}k (which does not fit the $\sqrt{n}$ distribution) are more blurred than for the membrane case, as expected for scattering from a less-sharp bubble edge.
Thus it seems that bubble shapes which reduce Friedel oscillations also effectively remove pseudomagnetic Landau effects due to the less uniform PMFs induced by their strain profiles.

Finally, we note that the square symbol energy peak at low energies in both bubble types is a state localized near the bubble edge and plotted in Figs. \ref{fig:results}f and \ref{fig:results}i.
It is not directly related to pseudomagnetic effects, but emerges due to the interface between the pristine graphene region outside the bubble and the strained, perturbed region within. 
The presence of localized states at this boundary acts somewhat like a potential, and induces the scattering which lies behind the Friedel oscillations in these bubbles.
We note that these states in the non-linear plate bubble are far less localized than their membrane bubble counterparts, due to an edge which is no longer as sharp.
This in turn leads to the smoothening and averaging out of the Friedel oscillations that we observed earlier for the non-linear plate bubbles.

\section{Conclusions}
We studied theoretically the local and averaged densities of states in gas-inflated graphene bubbles embedded in infinite graphene sheets by making use of the patched Green's function approach.
We determined that pseudo-Landau level features in sharp-edged bubbles may be hidden by interference effects due to electron scattering at the bubble edges.
Softer-edged bubbles were found to display weaker interference effects, however their shape profiles also resulted in pseudomagnetic field distributions unsuitable for pseudo-Landau level formation.
Our results suggest that it will be difficult to obtain reliable Landau level features in such gas inflated systems, unlike bubbles formed on substrates which often display the triaxial-type strain which is predicted to give a more appropriate pseudomagnetic field for Landau level formation.


\section*{References}
\providecommand{\newblock}{}

%

\end{document}